\documentclass[journal,letterpaper,twoside,twocolumn]{IEEEtran}
\usepackage{amsmath,amssymb,graphicx,psfrag,cite}
\usepackage{color}
\usepackage{balance} %

\newcommand{\eqlab}[2]{\begin{align} \label{#1} #2 \end{align}}
\newcommand{\eq}[1]{\begin{align*} #1 \end{align*}}

\newcommand{\bc}{{\boldsymbol{c}}}
\newcommand{\bw}{{\boldsymbol{w}}}
\newcommand{\bx}{{\boldsymbol{x}}}
\newcommand{\by}{{\boldsymbol{y}}}
\newcommand{\bX}{{\boldsymbol{X}}}
\newcommand{\bY}{{\boldsymbol{Y}}}
\newcommand{\E}{\mathbb{E}}
\newcommand{\R}{\mathbb{R}}

\newcommand{\amax}{{a_\mathrm{max}}}

\newcommand{\todo}[1]{}

\newtheorem{theorem}{Theorem}

\renewcommand{\markboth}[1]
  {\renewcommand{\leftmark}{#1}\renewcommand{\rightmark}{#1}}
\title{The Channel Capacity Increases with Power}
\author{Erik Agrell
\thanks{This is the third and last version of ``The channel capacity increases with power.'' Future improvements, including results for multiuser channels, will be presented under the title ``On monotonic capacity--cost functions,'' http://arxiv.org/abs/1209.2820.

This work was supported in part by the Swedish Foundation for Strategic Research (SSF) under grant RE07-0026 and the Swedish Research Council (VR) under grant 2007-6223. E.~Agrell is with the Dept.~of Signals and Systems, Chalmers Univ.~of Technology, SE-41296 G\"oteborg, Sweden (e-mail: agrell@chalmers.se).}}

\begin{document}
\maketitle

\begin{abstract}
It is proved that for memoryless vector channels, maximizing the mutual information over all source distributions with a certain average power or over the larger set of source distributions with upperbounded average power yields the same channel capacity in both cases. Hence, the channel capacity cannot decrease with increasing average transmitted power, not even for channels with severe nonlinear distortion.
\end{abstract}


\section{Introduction} %

\todo{
search for xx and \todo
search for "capacity"
place figures
balance columns on last page
read through it all
proofread references
}

\IEEEPARstart{I}{n the most} cited paper in the history of information theory \cite{shannon48}, Shannon proved that with adequate coding, reliable communication is possible over a noisy channel, as long as the rate does not exceed a certain threshold, called the \emph{channel capacity.} He provided in 1948 a mathematical expression for the channel capacity of any channel, based on its statistical properties. The expression is given as the supremum over all possible source distributions of a quantity later called the \emph{mutual information} \cite{shannon56, kolmogorov56}. The channel capacity is often studied as a function of the average transmitted power. This function is obtained by optimizing the mutual information over all source distributions whose average second moment is either \emph{equal} to the given power or \emph{upperbounded} by this power---the convention differs between disciplines. We will return to the distinction between the two definitions at the end of this section.

For linear channels with additive, signal-independent noise, the channel capacity is an increasing function of the transmitted power. The most well-known example is the \emph{additive white Gaussian noise} (AWGN) channel, for which the channel capacity is known exactly \cite[Sec.~24]{shannon48}, \cite[Ch.~9]{cover06}. In recent years, the problem of calculating or estimating the channel capacity of more complicated channels has received a lot of attention (see surveys in \cite{katz04, kahn04, essiambre10}).  Due to the absence of exact analytical solutions and the computational intractability of optimizing over all possible source distributions, most investigations of the channel capacity of non-AWGN channels rely on bounding techniques and asymptotic analysis.

If only \emph{noncoherent detection} is available at the receiver, the channel capacity can be analyzed by including a magnitude operation at the output of a discrete-time complex AWGN channel. The channel capacity is in this case not known exactly, but it increases logarithmically with transmitted power as approximately half the regular AWGN channel capacity \cite{blachman53}, \cite[Sec.~11.2]{ho05}. The same behavior has been shown for the \emph{phase-noise channel,} in which the transmitted signal is subject to a uniformly random phase shift before the Gaussian noise is added \cite{lapidoth02, katz04}; indeed, according to \cite{katz04}, these two channel models are equivalent in terms of channel capacity.

For the \emph{Rayleigh-fading} channel, the channel capacity increases logarithmically with power, with just an asymptotic offset to the AWGN channel capacity, if the receiver has full channel state information \cite{ericsson70, lapidoth03}. The increase is doubly logarithmic if no channel state information is available \cite{taricco97, abou-faycal01, lapidoth03}. The results have been extended to other wireless channel models, including Rician fading, systems with transmitter-side channel state information, and multiple-antenna channels, see \cite{lapidoth03, taricco11}, \cite[Sec.~4.2--4.3, 10.3, 14.5--14.7]{goldsmith05} and references therein. In all these cases, the channel capacity is an increasing function of the transmitted power.

Of particular interest for this paper is the type of nonlinear distortion encountered in fiber-optical communications \cite{chraplyvy90}, \cite[Sec.~7.2]{agrawal08book}. The impact of this nonlinear distortion increases dramatically with the transmitted power, to the extent that communication becomes virtually impossible if the instantaneous power is high enough \cite{demir07}, \cite[Ch.~9]{agrawal10}. This phenomenon is well known from experiments and simulations. Thus one might expect that the mutual information and channel capacity would approach zero at sufficiently high power.

If the mutual information is computed for a given source distribution, or optimized over a subset of all possible source distribution, a lower bound on the channel capacity is obtained. Numerous such lower bounds have been derived for nonlinear fiber-optical channels. The earliest bounds for optical channel capacity assumed a Gaussian source density, for which the mutual information can be calculated or lowerbounded analytically \cite{mitra01, green02, wegener04, kahn04, ellis10, ellis11, killey11}. If the source is constrained to a ring with constant amplitude, another bound is obtained, which is stronger under some conditions \cite{ho02}, \cite[Sec.~11.4]{ho05}. In recent studies, the mutual information has been optimized numerically for concentric multiring constellations \cite{essiambre08, freckmann09, essiambre10, goebel11, djordjevic10, smith10, djordjevic11}, which yields other bounds on the channel capacity. Interestingly, all these lower bounds show the same general trend: As the average power (or signal-to-noise ratio) increases, they increase towards a peak, and then they decrease again towards zero as the power is further increased, similarly to most of the curves in Figs.~\ref{fig:ixy}--\ref{fig:constrained}. This is not unexpected since, as mentioned above, the severity of the nonlinear distortion increases with power. In contrast, \cite{turitsyn03} and \cite{taghavi06} indicate that the channel capacity may increase monotonically for certain nonlinear optical channels.

Some of the lower bounds mentioned above are derived for single-user systems and others for multiuser systems. In optical communications, a multiuser system is sometimes modeled as a channel where the distortion, representing interference from other users, changes depending on the transmitted power---i.e., a source-dependent channel. Such channels will not be considered in this paper. We focus on single-user systems, where the channel output can be represented by a single, fixed distribution conditioned on the input, which is the scenario considered in Shannon's original work \cite{shannon48}.

Apart from lower bounds, not much is known about the channel capacity of nonlinear fiber-optical channels. To the author's knowledge, no generally accepted expressions are available for the \emph{exact} channel capacity\footnote{Exact expressions were proposed in \cite{narimanov02} and in \cite{tang01, tang06}, but their validity was questioned in \cite{kahn04} andÊ\cite{turitsyn03}, resp.} and no \emph{upper} bounds are known, apart from the standard channel capacity for the linear AWGN channel, which neglects all nonlinear distortion. Unfortunately, the numerous articles about lower bounds have often been cited in terms of just channel capacity (or capacity, spectral efficiency, information spectral density, etc.), without mentioning that the cited results are bounds. Therefore, there is a wide-spread belief in the optical community that the channel capacity of channels with strong nonlinear distortion increases with power to a certain maximum value and then decreases again towards zero.
 
In this paper, we prove rigorously that the channel capacity is an increasing (but not necessarily strictly increasing) function of the average power, thus disproving the standard belief of a peaky behavior in the single-user case. There is no contradiction between this result and the numerous decreasing lower bounds referenced above, but we do recommend some caution in drawing conclusions about the true channel capacity from the behavior of its lower bounds alone. It is beyond doubt that the mutual information can decrease with power, as do many lower bounds on the channel capacity, but not channel capacity in the Shannon sense. This fundamental theorem is proved for a general continuous-input, continuous-output channel, not confined to any particular channel model or application.

The theorem holds regardless of whether the given power level is interpreted as the exact second moment of the source or an upper bound thereof. The proof is developed assuming the former definition, and it is trivial for the latter. An interesting consequence of the increasing channel capacity is that the two definitions of channel capacity are fully equivalent.

\section{Mutual Information, Constrained Capacity, and Channel Capacity} \label{sec:def}%

For any random vectors $\bY$ and $\bX$ of length $n$, the (differential) \emph{entropy} $h(\bY)$ and \emph{conditional entropy} $h(\bY|\bX)$ are defined as \cite[Sec.~20]{shannon48}
\eqlab{eq:hy}{
  h(\bY) &\triangleq -\int_{\R^n} f_\bY(\by) \log f_\bY(\by) d\by, \\
  h(\bY|\bX) &\triangleq -\int_{\R^n} \int_{\R^n} f_{\bX\bY}(\bx,\by) \log f_{\bY|\bX}(\by|\bx) d\bx d\by, \label{eq:hyx}
}
where $f_\bY(\by)$ denotes the probability density function (pdf) of $\bY$, $f_{\bY|\bX}(\by|\bx)$ is the conditional pdf of $\bY$ given $\bX$, and $f_{\bX\bY}(\bx,\by)$ is the joint pdf of $\bX$ and $\bY$, assuming that these pdfs all exist. All logarithms are to base 2 and $0 \log 0$ should be interpreted as 0.\footnote{This convention can be made rigorous by confining the integrals to the support of the involved random variables.}
The \emph{mutual information} in bits/symbol between $\bX$ and $\bY$ is defined as \cite{shannon56, kolmogorov56}, \cite[Sec.~8.5]{cover06}\footnote{If any of the above mentioned pdfs do not exist, the mutual information is still well defined in terms of finite partitions \cite[eq.~(8.54)]{cover06}.}
\eqlab{eq:ixy}{
  I(\bX;\bY) \triangleq h(\bY) - h(\bY|\bX)
.}

Let $\bX$ and $\bY$ represent the input and output, resp., of a communication channel. The joint distribution $f_{\bX\bY}(\bx,\by)$ can be factorized as $f_{\bX\bY}(\bx,\by) = f_\bX(\bx) f_{\bY|\bX}(\by|\bx)$, where $f_\bX$ represents the source and $f_{\bY|\bX}$ represents the channel. The source distribution $f_\bX$ is usually chosen to match a certain channel $f_{\bY|\bX}$, but the converse is not realistic for single-user channels: the channel should be represented by the same function $f_{\bY|\bX}(\by|\bx)$ regardless of the source. For any given source distribution $f_\bX(\bx)$, the mutual information can be calculated from \eqref{eq:hy}--\eqref{eq:ixy}, where $f_\bY(\by)$ is obtained by marginalizing $f_{\bX\bY}(\bx,\by)$ over $\bx$.

The supremum of these mutual informations, over all possible source distributions $f_\bX$, is the \emph{channel capacity} \cite[Sec.~23]{shannon48}, \cite{verdu94}, \cite[p.~274]{cover06}. In this paper, we study the channel capacity $C$ as a function of the average transmitted power $P\ge 0$, defined as the second moment of the source distribution. This is a special case of a \emph{capacity--cost function}, where cost is the average transmitted power. The function can be defined in two, subtly different, ways, depending on whether the power (cost) is \emph{upperbounded} by $P$ or \emph{exactly} $P$. In the first case, which is most common in classical information theory \cite[Ch.~7]{gallager68}, \cite{verdu90}, \cite[Ch.~9]{cover06}, the channel capacity is defined as
\eqlab{eq:cprime}{
  C'(P) \triangleq \sup_{f_\bX \in \Omega'(P)} I(\bX;\bY)
,}
where $\Omega'(P)$ is the set of all distributions\footnote{The distribution may be discrete, continuous, or mixed.}
over $\R^n$ such that $\E[\|\bX\|^2] \le P$. In the second case, which is prevalent in optical information theory \cite{mitra01,wegener04}, the channel capacity is
\eqlab{eq:c}{
  C(P) \triangleq \sup_{f_\bX \in \Omega(P)} I(\bX;\bY)
,}
where $\Omega(P)$ is the set of all distributions over $\R^n$ such that $\E[\|\bX\|^2] = P$. In both cases, $f_{\bY|\bX}(\by|\bx)$ is assumed to be independent of $P$; i.e., the channel statistics do not depend on the source statistics.

Since $\Omega'(P) \supseteq \Omega(P)$, $C'(P) \ge C(P)$ for all $P$ and all channels. Furthermore, since $\Omega'(P_2) \supseteq \Omega'(P_1)$ for $P_1 \ge P_2$, $C'(P)$ is a nondecreasing function of $P$ \cite[p.~3-21]{elgamal10}. In this paper, ``channel capacity'' refers to $C(P)$ unless otherwise stated. However, the two definitions are in fact equivalent, as will be shown in Theorem \ref{th:equivalent}.

If the optimization of $I(\bX;\bY)$ is instead done over a subset of $\Omega(P)$ (or $\Omega'(P)$), a \emph{constrained capacity} is obtained. Many versions of constrained capacity have been studied in the past, such as confining $\bX$ to a certain range or to a certain discrete constellation.

To summarize the terminology used in this paper, we will use ``mutual information'' when no optimization is carried out, ``constrained capacity'' when the optimization is over some, but not all, possible source distributions, and ``channel capacity'' when the optimization is over all possible source distributions. Thus, the mutual information between input and output is a property of the channel and the source, the constrained capacity is a property of the channel and the source constraints, and the channel capacity is a property of the channel alone. To avoid confusion, we will not use just ``capacity'' in this paper.

\section{The Law of Monotonic Channel Capacity} \label{sec:main} %

We are now ready to state the main result. It is given in terms of an arbitrary discrete-time, memoryless, vector channel. This general channel model includes the discrete-time channel with memory, if the dimension (block length) is chosen large enough \cite{verdu94}, and also the continuous-time bandlimited channel, because bandlimited waveforms can without loss be represented by a vector of its samples \cite[Sec.~23]{shannon48}.

\begin{theorem}[Law of monotonic channel capacity] \label{th:main}
$C(P)$ is a nondecreasing function of $P$ for any channel $f_{\bY|\bX}(\by|\bx)$.
\end{theorem}

\begin{IEEEproof}
Let $f_{\bX'} \in \Omega(P')$ be a capacity-achieving distribution at any power $P'\ge 0$. We will show that $C(P) \ge C(P')$ for any $P \ge P'$.

For any given $P\ge P'$ and $0<\epsilon\le 1$, let 
\eq{
P'' \triangleq P'+\frac{P-P'}{\epsilon}
}
and let $f_{\bX''}$ be any distribution over $\R^n$ with power $\E[\|\bX''\|^2] = P''$. We define a time-sharing random vector $\bX$ given an auxiliary binary random variable $Q$ such that
\eqlab{eq:fx2}{
  \bX \triangleq \begin{cases}
    \bX', & Q = 0, \\
    \bX'', & Q = 1,
  \end{cases}
}
where $\Pr\{Q = 1\} = \epsilon$. The power of $\bX$ is
\eq{
  \E[\|\bX\|^2] &= (1-\epsilon)\E[\|\bX'\|^2]+\epsilon \E[\|\bX''\|^2] \\
  & = (1-\epsilon)P' + \epsilon P'' \\
  & = P
  .}

Because $Q\rightarrow\bX\rightarrow\bY$ is a Markov chain, the mutual information can be bounded as
\eqlab{eq:i-rescaled0}{
I(\bX,\bY) &\ge I(\bX;\bY | Q) \nonumber\\
  &= (1-\epsilon) I(\bX;\bY | Q=0) + \epsilon I(\bX;\bY | Q=1) \nonumber\\
  &\ge (1-\epsilon) I(\bX;\bY | Q=0) \nonumber\\
  &= (1-\epsilon) I(\bX';\bY') \nonumber\\
  &= (1-\epsilon) C(P')
. }
Thus
\eqlab{eq:lowerbound}{
C(P) &= \sup_{f_{\bX}\in\Omega(P)} I(\bX;\bY) \nonumber\\
  &\ge \sup_{0<\epsilon\le 1} (1-\epsilon) C(P') \nonumber\\
  &= C(P')
  ,}
which completes the proof.
\end{IEEEproof}

A practical interpretation of the theorem is that it is possible to waste power without sacrificing channel capacity. Even though this is a huge improvement over previous results, where the channel capacity was believed to decay to zero, a system designer would not be too excited over the possibility to waste power without gaining anything. However, \eqref{eq:lowerbound} is only a lower bound, obtained when the source distribution has the special form \eqref{eq:fx2}. This form was chosen because of its general applicability to any channel $f_{\bY|\bX}(\by|\bx)$, but it is also not optimized for any channel. If \eqref{eq:fx2} would be replaced by a distribution optimized for a given channel and transmitted power, $I(\bX_2;\bY_2)$ may increase, and a tighter lower bound than \eqref{eq:lowerbound} would be obtained. This would make the channel capacity strictly increasing with power, which in addition to its theoretical significance may have practical implications for the design of communication systems operating in the nonlinear regime.

An immediate consequence of Theorem \ref{th:main}, given by the next theorem, is that the power-limited channel capacity $C'(P)$ is achieved by a source distribution $f_\bX$ for which the second moment equals the maximum allowed value $P$. This means that the two definitions \eqref{eq:cprime} and \eqref{eq:c} are equivalent.

\begin{theorem}\label{th:equivalent}
For any channel and any $P$, $C'(P) = C(P)$.
\end{theorem}

\begin{IEEEproof}
The definition \eqref{eq:cprime} can be written as
\eq{
  C'(P) = \sup_{P_1 \le P} C(P_1)
,}
which by Theorem \ref{th:main} is equal to $C(P)$.
\end{IEEEproof}

\section{Numerical Examples} %

In this section, examples will be presented for mutual information, constrained capacity, and channel capacity as functions of the average transmitted power, where the mutual information and constrained capacity have peaks but the channel capacity, as predicted by Theorem \ref{th:main}, is nondecreasing.

\subsection{A Nonlinear Channel}%

We consider a very simple channel with nonlinear distortion and noise, represented as
\eqlab{eq:channel}{
Y = a(X)+Z
,}
where $X$ and $Y$ are the real, scalar input and output of the channel, resp., $a(\cdot)$ is a deterministic, scalar function and $Z$ is white Gaussian noise with zero mean and variance $\sigma_Z^2$. For a given channel input $x$, the distribution of the channel output is represented by the conditional probability density function (pdf)
\eqlab{eq:fyx}{
f_{Y|X}(y|x) = \frac{1}{\sigma_Z} f_G \left( \frac{y-a(x)}{\sigma_Z} \right)
,}
where $f_G(x) \triangleq (1/\sqrt{2\pi}) \exp(-x^2/2)$ is the zero-mean, unit-variance Gaussian pdf.
The corresponding conditional entropy for a given source pdf $f_X(x)$ is given by \eqref{eq:hyx} as
\eq{
h(Y|X) = -\int_{-\infty}^\infty f_X(x) \int_{-\infty}^\infty f_{Y|X}(y|x) \log f_{Y|X}(y|x) dy dx
.}
Since $f_{Y|X}(y|x)$ is Gaussian for a given $x$, the inner integral is \cite[Sec.~20]{shannon48}, \cite[Sec.~8.1]{cover06}
\eq{
\int_{-\infty}^\infty f_{Y|X}(y|x) \log f_{Y|X}(y|x) dy = -\frac{1}{2}\log2\pi e \sigma_Z^2
}
independently of $x$ and hence
\eqlab{eq:hyx3}{
h(Y|X) = \frac{1}{2}\log2\pi e \sigma_Z^2
.}
The mutual information for this channel and any source pdf $f_X(x)$ is $I(X;Y) = h(Y)-h(Y|X)$, where
\eqlab{eq:hy-scalar}{
h(Y) &= -\int_{-\infty}^\infty f_Y(y) \log f_Y(y) dy, \\
f_Y(y) &= \int_{-\infty}^\infty f_X(x) f_{Y|X}(y|x) dx \nonumber
,}
$f_{Y|X}(y|x)$ is given by \eqref{eq:fyx}, and $h(Y|X)$ is given by \eqref{eq:hyx3}.

\psfrag{ax}{\footnotesize $a(x)$}
\psfrag{x}{\footnotesize $x$}

\begin{figure}
\begin{center}
\includegraphics[width=\columnwidth]{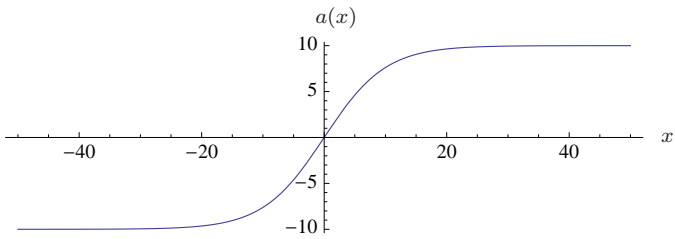}
\caption{A simple example of nonlinear distortion, given by \eqref{eq:tanh} for $\amax=10$. The channel is essentially linear for small $|x|$ and binary for large $|x|$.}
\label{fig:tanh}
\end{center}
\end{figure}

As an example, we select $a(x)$ in \eqref{eq:channel} as a smooth clipping function
\eqlab{eq:tanh}{
a(x) = \amax \tanh \left( \frac{x}{\amax} \right)
,}
where $\amax>0$ sets an upper bound on the output. This channel is chosen for its simplicity and because its characteristics serves to illustrate the Law of monotonic channel capacity (Theorem \ref{th:main}), not for its resemblance to any particular physical system. If the instantaneous channel input $X$ has a sufficiently high magnitude, the channel is essentially binary. For $X$ close to zero, on the other hand, the channel approaches a linear AWGN channel.

The channel parameters are $\amax=10$ and $\sigma_Z=1$ throughout this paper. The function $a(x)$ in \eqref{eq:tanh}, which represents the nonlinear part of the channel \eqref{eq:channel}, is shown in Fig.~\ref{fig:tanh}.

\begin{figure}
\begin{center}
\psfrag{x}{\footnotesize $P$}
\psfrag{y}{\footnotesize $I(X;Y)$}
\psfrag{99}[][]{\scriptsize $0.1$}
\psfrag{0}[][]{\scriptsize $1$}
\psfrag{10}[][]{\scriptsize $10$}
\psfrag{20}[][]{\scriptsize $100$}
\psfrag{30}[][]{\scriptsize $1000$}
\psfrag{40}[][]{\scriptsize $10^4$}
\psfrag{50}[][]{\scriptsize $10^5$}
\includegraphics[width=\columnwidth]{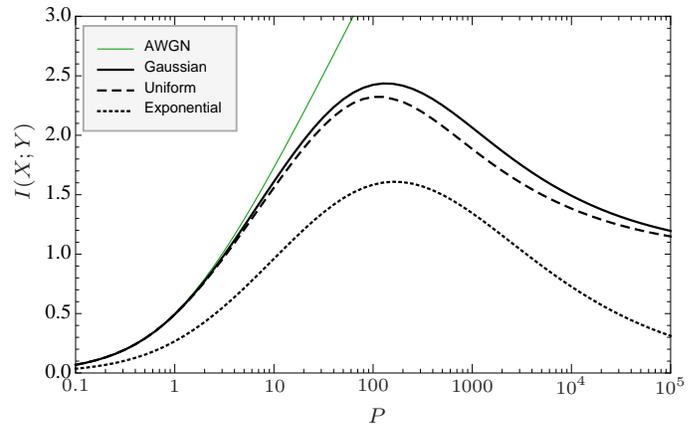}
\caption{Mutual information for the nonlinear channel in \eqref{eq:channel} with $\amax=10$ and $\sigma_Z=1$, when the source pdf is Gaussian, uniform, and exponential. The AWGN channel capacity is included for reference.}
\label{fig:ixy}
\end{center}
\end{figure}

\subsection{Mutual Information}%

The mutual information $I(X;Y)$ is evaluated by numerical integration, as a function of the transmitted power $P$. No optimization over pdfs is carried out. The source pdf $f_X(x)$ is constructed from a given unit-power distribution $g(x)$, rescaled to the desired power $P$ as $f_X(x) = \beta g(\beta x)$, where $\beta=1/\sqrt{P}$. The results are presented in Fig.~\ref{fig:ixy} for three continuous source pdfs $f_X(x)$: zero-mean Gaussian, zero-mean uniform, and single-sided exponential, defined as, respectively,
\eq{
f_{X_1}(x) &= \frac{1}{\sqrt{P}} f_G\left(\frac{x}{\sqrt{P}}\right), \\
f_{X_2}(x) &= \begin{cases}
  \frac{1}{2\sqrt{3P}}, & -\sqrt{3P} \le x \le \sqrt{3P}, \\
  0, & \text{elsewhere},
\end{cases} \\
f_{X_3}(x) &= \begin{cases}
  \sqrt{\frac{2}{P}}e^{-x \sqrt{2/P}}, & x \ge 0, \\
  0, & x < 0.
\end{cases}
}

At asymptotically low power $P$, the channel is effectively an AWGN channel. In this case, the mutual information is governed by the mean value of the source distribution, according to \cite[Th.~7]{agrell11}. All zero-mean sources achieve approximately the same mutual information, which approaches the AWGN channel capacity. The asymptotic mutual information for the exponential distribution, whose mean is $\sqrt{P/2}$, is half that achieved by zero-mean distributions.

The mutual information curves for all three source pdfs reach a peak around $P=100$, when a large portion of the source samples still fall in the linear regime of the channel. When the average power $P$ is further increased, the mutual information decreases towards a value slightly less than 1 asymptotically for the zero-mean sources and 0 for the exponential source. The asymptotes are explained by the fact that at high enough power, almost all source samples fall in the nonlinear regime, where the channel behaves as a 1-bit noisy quantizer.

\begin{figure}
\begin{center}
\psfrag{x}{\footnotesize $P$}
\psfrag{y}{\footnotesize $I(X;Y)$}
\psfrag{99}[][]{\scriptsize $0.1$}
\psfrag{0}[][]{\scriptsize $1$}
\psfrag{10}[][]{\scriptsize $10$}
\psfrag{20}[][]{\scriptsize $100$}
\psfrag{30}[][]{\scriptsize $1000$}
\psfrag{40}[][]{\scriptsize $10^4$}
\psfrag{50}[][]{\scriptsize $10^5$}
\includegraphics[width=\columnwidth]{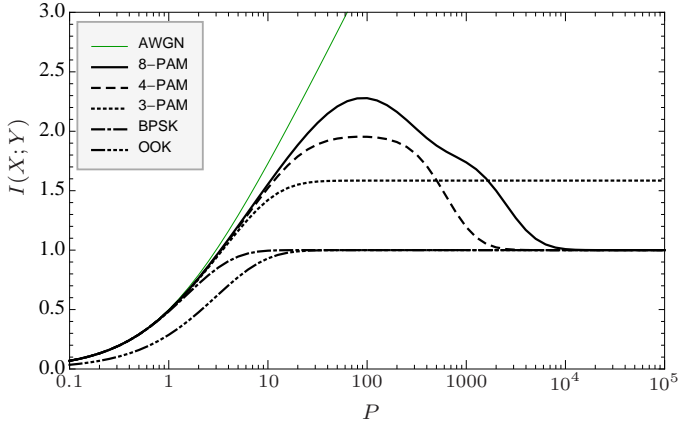}
\caption{Mutual information for the same channel, when the source follows various discrete distributions. The source probabilities are uniform.}
\label{fig:ixy2}
\end{center}
\end{figure}

Similar results for various discrete source distributions are shown in Fig.~\ref{fig:ixy2}. The studied one-dimensional constellations are on--off keying (OOK), binary phase-shift keying (BPSK), and $M$-ary pulse amplitude modulation ($M$-PAM). The constellation points are equally spaced and the source samples $X$ are chosen uniformly from these constellations. The mutual information for $M$-PAM constellations with $M\ge 4$ exhibits the same kind of peak as the continuous distributions in Fig.~\ref{fig:ixy}; indeed, a uniform distribution over equally spaced $M$-PAM approaches the continuous uniform distribution as $M \rightarrow \infty$.

Similarly to the continuous case, the zero-mean discrete sources approach the AWGN channel capacity as $P\rightarrow 0$. Half this channel capacity is achieved by the OOK source, which has the same mean value $\sqrt{P/2}$ as the exponential source above. The asymptotics when $P\rightarrow\infty$ depends on whether $M$ is even or odd. For any even $M$, the channel again acts like a 1-bit quantizer and the asymptotic mutual information is slightly less than 1. For odd $M$, however, here exemplified by $3$-PAM, there is a nonzero probability mass at $X=0$, which means that the possible outputs are not only $Y=\pm \amax+Z$ but also $Y=0+Z$. Hence, the channel asymptotically approaches a ternary-output noisy channel, whose mutual information is upperbounded by $\log 3 = 1.58$.

To summarize, this particular channel has the property that the mutual information for any source distribution approaches a limit as $P\rightarrow\infty$, and this limit is upperbounded by $\log 3$. It might seem tempting to conclude that the channel capacity, which is the supremum of all mutual information curves, would behave similarly. However, as we shall see in Section \ref{sec:channelcap}, this conclusion is not correct, because the limit of a supremum is in general not equal to the supremum of a limit. Specifically, the asymptotical channel capacity is $\lim_{P\rightarrow\infty} C(P) = \lim_{P\rightarrow\infty} \sup_{g} I(X;Y)$, which is not equal to $\sup_{g} \lim_{P\rightarrow\infty} I(X;Y) \le \log 3$.

\subsection{Constrained Capacity} \label{sec:constrainedcap} %

\begin{figure}
\begin{center}
\psfrag{x}{\footnotesize $P$}
\psfrag{99}[][]{\scriptsize $0.1$}
\psfrag{0}[][]{\scriptsize $1$}
\psfrag{10}[][]{\scriptsize $10$}
\psfrag{20}[][]{\scriptsize $100$}
\psfrag{30}[][]{\scriptsize $1000$}
\psfrag{40}[][]{\scriptsize $10^4$}
\psfrag{50}[][]{\scriptsize $10^5$}
\psfrag{y}{\footnotesize $\sup I(X;Y)$}
\includegraphics[width=\columnwidth]{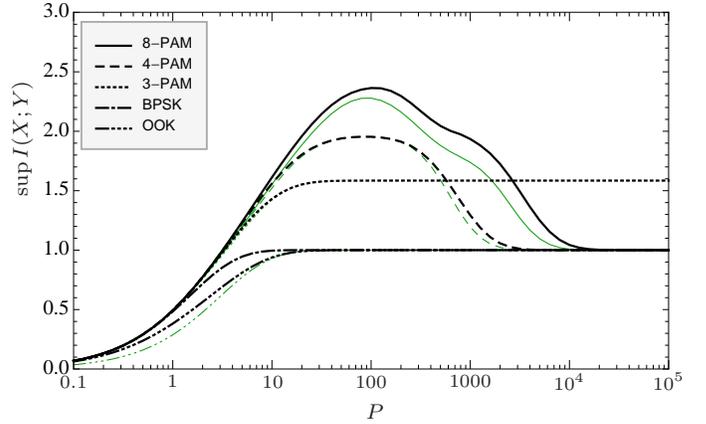}
\caption{Constrained capacities for the same channel, where the source is constrained to a given constellation but the source probabilities are optimally chosen for each $P$. The colored curves indicate the nonoptimized mutual informations from Fig.~\ref{fig:ixy2} (uniform probabilities).}
\label{fig:constrained}
\end{center}
\end{figure}

The standard method to calculate the channel capacity of a discrete memoryless channel is by the \emph{Blahut--Arimoto algorithm} \cite[Sec.~10.8]{cover06}, \cite[Ch.~9]{yeung08}. It has been extended to continuous-input, continuous-output channels in \cite{chang88,dauwels05}. Our approach is most similar to \cite{dauwels05}, in which pdfs are represented by lists of samples, so-called \emph{particles.} We consider a source pdf of the form
\eqlab{eq:particle}{
f_X(x) = \sum_{i=1}^N w_i \delta(x-c_i)
,}
where $\delta(\cdot)$ is the Dirac delta function, $N$ is the number of samples, $\bc=(c_1,\ldots,c_N)$ are the samples, and $\bw=(w_1,\ldots,w_N)$ are the probabilities, or weights, associated with each sample. If $N$ is large enough, any pdf can be represented in the form \eqref{eq:particle} with arbitrarily small error. With this representation,
\eq{
f_Y(y) = \sum_{i=1}^N \frac{w_i}{\sigma_Z} f_G \left( \frac{y-a(c_i)}{\sigma_Z} \right)
,}
which when substituted in \eqref{eq:hy-scalar} yields $h(Y)$, and thereby $I(X;Y)$, by numerical integration.

The objective for the optimization is to maximize the Lagrangian function
\begin{multline*}
L(\bc,\bw,\lambda_1,\lambda_2)
  \triangleq h(Y)+\lambda_1\left( \sum_{i=1}^N w_i - 1 \right) \\
  \qquad+\lambda_2\left( \sum_{i=1}^N w_i c_i^2 - P \right)
,
\end{multline*}
where the Lagrange multipliers $\lambda_1$ and $\lambda_2$ are determined to maintain the constraints $\sum_i w_i = 1$ and $\sum_i w_i c_i^2 = P$ during the optimization process. The gradients of $L$ with respect to $\bc$ and $\bw$ are calculated, and a steepest descent algorithm (or more accurately, ``steepest ascent'') is applied to maximize $L$. In each iteration, a step is taken in the direction of either of the two gradients.\footnote{Moving in the direction of the joint gradient turned out to be less efficient, because for small and large $P$, the numerical values of $\bc$ and $\bw$ are not of the same order of magnitude.} The step size is determined using the \emph{golden section method} \cite[pp.~271--273]{lundgren10}. Constrained capacities were obtained by including additional constraints on $\bc$ and/or $\bw$. Several initial values $(\bc,\bw)$ were tried, and $N$ was increased until convergence.

The topography of $L$ as a function of $\bc$ and $\bw$ turned out to include vast flat fields, where a small step has little influence on $L$. This made the optimization numerically challenging. No suboptimal local maxima were found for the studied channel and constraints, although for nonlinear channels in general, the mutual information as a function of the source distribution may have multiple maxima.\footnote{An exception occurs when the constellation points $\bc$ are fixed and the only constraint is $\sum w_i=1$. In this special case, the mutual information is a concave function of $\bw$ for any channel \cite[pp.~33, 191]{cover06} and there is thus a unique maximum.}

\begin{figure}
\begin{center}
\psfrag{x}{\footnotesize $P$}
\psfrag{99}[][]{\scriptsize $0.1$}
\psfrag{0}[][]{\scriptsize $1$}
\psfrag{10}[][]{\scriptsize $10$}
\psfrag{20}[][]{\scriptsize $100$}
\psfrag{30}[][]{\scriptsize $1000$}
\psfrag{40}[][]{\scriptsize $10^4$}
\psfrag{50}[][]{\scriptsize $10^5$}
\psfrag{y}{\footnotesize $C(P)$}
\includegraphics[width=\columnwidth]{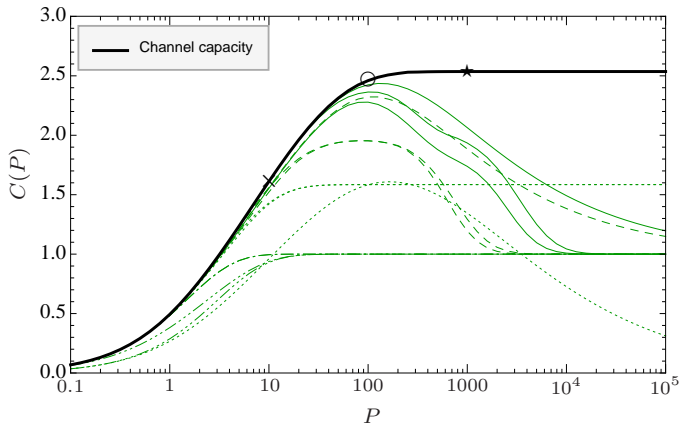}
\caption{Channel capacity for the same channel. All mutual information and constrained capacity curves from Figs.~\ref{fig:ixy}--\ref{fig:constrained} are included for reference (colored). Even though most mutual information and constrained capacity curves decrease, the channel capacity does not. The three markers refer to distributions in Fig.~\ref{fig:pmfs}.}
\label{fig:channelcap}
\end{center}
\end{figure}

Using this optimization technique, some constrained capacities are computed. Specifically, we investigate how much the mutual information curves in Fig.~\ref{fig:ixy2} can be improved if the source samples $X$ are chosen from the constellation points $\bc$ with unequal probabilities $\bw$, so-called \emph{probabilistic shaping}. The constellations are the same as before, equally spaced OOK, BPSK, and $M$-PAM, but the probabilities of each constellation point is allowed to vary. For each power $P$, the mutual information is maximized over all probabilities. The constellation is scaled to meet the power requirement but otherwise not changed.

\begin{figure*}
\begin{center}
\psfrag{x}{\footnotesize $x$}
\psfrag{amax}{\footnotesize $\amax$}
\psfrag{pow10}{\small $P=10$}
\psfrag{pow100}{\small $P=100$}
\psfrag{pow1000}{\small $P=1000$}
\includegraphics[width=\textwidth]{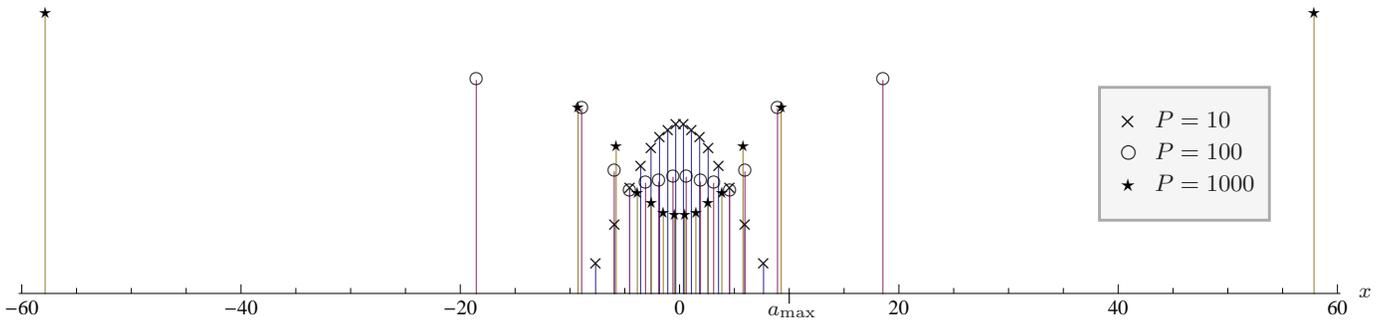}
\caption{Discrete approximations of the capacity-achieving source distributions for $P=10$, $100$, and $1000$.}
\label{fig:pmfs}
\end{center}
\end{figure*}

The results are shown in Fig.~\ref{fig:constrained} for the same channel as before (\eqref{eq:channel} with \eqref{eq:tanh} and parameters $\amax=10$ and $\sigma_Z=1$). The BPSK performance offers no improvement over the mutual information of uniform BPSK in Fig.~\ref{fig:ixy}, because equal probabilities turn out to be optimal for all $P$. However, the constrained capacity of OOK with optimal probabilistic shaping is about twice the mutual information of uniform OOK at low $P$. The improvements for $3$- and $4$-PAM are marginal, whereas the performance of $8$-PAM is significantly improved for medium to high power, and its peak increases from 2.28 to 2.37 bits/symbol. The general trends, however, are the same as for the mutual information in Fig.~\ref{fig:ixy2}: The constrained capacity for any probabilistically shaped $M$-PAM system with $M \ge 4$ displays a prominent peak around $P=100$, after which the constrained capacity decreases again towards the same asymptotes as in the uniform case.

Obviously, there exist many other types of source constraints. Some of these have constrained capacities similar to those of the probabilistically shaped discrete constellations shown in Fig.~\ref{fig:constrained}, with a peak at a finite power and a relatively weak asymptotic performance, but other classes of sources can be conceived that are better suited to this nonlinear channel at high transmitted power. However, instead of designing further constrained sources, we will now proceed to study the channel capacity, which is the main concern of this paper.

\subsection{Channel Capacity} \label{sec:channelcap} %

By optimizing the mutual information over unconstrained source distributions $\Omega(P)$, according to the method outlined in Sec.~\ref{sec:constrainedcap}, we obtain the channel capacity \eqref{eq:c}. As mentioned in Sec.~\ref{sec:def}, the channel capacity is a property of the channel alone, not the source, so there exists just one channel capacity curve for a given channel.

This channel capacity is shown in Fig.~\ref{fig:channelcap} for the studied channel. As promised by the Law of monotonic channel capacity (Theorem \ref{th:main}), the curve does not have any peak at a finite $P$, which characterizes most mutual information and constrained capacity curves. The channel capacity follows the mutual information of the Gaussian distribution closely until around $P=100$. However, while the Gaussian case attains its maximum mutual information $I(X;Y) = 2.44$ bits/symbol at $P=130$ and then begins to decrease, the channel capacity continues to increase towards its asymptote $\lim_{P\rightarrow \infty} C(P) = 2.54$ bits/symbol.

This asymptotical channel capacity can be explained as follows. Define the random variable $A\triangleq a(X)$. Since $a(\cdot)$ is a continuous, strictly increasing function, there is a one-to-one mapping between $X \in (-\infty,\infty)$ and $A \in (-\amax,\amax)$. Thus $I(X;Y) = I(A;Y)$, where $Y=A+Z$. This represents a standard discrete-time AWGN channel whose input $A$ is subject to a peak power constraint. The constrained capacity of a peak-power-limited AWGN channel was bounded already in \cite[Sec.~25]{shannon48} and computed numerically in \cite{smith71}, where it was also shown that the capacity-achieving distribution is discrete. The asymptote in Fig.~\ref{fig:channelcap}, which is $2.54$ bits/symbol or, equivalently, $1.76$ nats/symbol, agrees perfectly with the constrained capacity in \cite[Fig.~2]{smith71} for $\amax/\sigma_Z = 10$.

Some of the (almost) capacity-achieving source distributions are shown in Fig.~\ref{fig:pmfs}, numerically optimized as described in Sec.~\ref{sec:constrainedcap}. As mentioned, the topography of $L$ as a function of the source parameters for a given $P$ includes vast, almost flat, fields, where many source distributions yield the same mutual information, within a numerical precision of 2--3 decimals. For $P=10$, the optimized discrete source is essentially a nonuniformly sampled Gaussian pdf, and the obtained channel capacity, 1.61, has the same value as the mutual information of a continuous Gaussian pdf, shown in Fig.~\ref{fig:ixy}. For $P=100$ and $1000$, the distribution is more uniform in the range where the channel behaves more or less linearly, which for this channel is approximately at $-\amax/2<x<\amax/2$, with some high-power outliers in the nonlinear range $|x| > \amax$. In all cases, increasing the number of particles $N$ from what is shown in Fig.~\ref{fig:pmfs} does not increase the mutual information significantly, from which we infer that these discrete sources perform practically as well as the best discrete or continuous sources for this channel.

Although the capacity-achieving distributions would look quite different for other types of nonlinear channels, a general observation can be made from Fig.~\ref{fig:pmfs}: Even at high average power, the source should generate samples with moderate power, for which the channel is good, most of the time. The high average power is achieved by some samples having a very large power but small probability. The two-part distribution \eqref{eq:fx2}, which was used to prove Theorem \ref{th:main}, can be seen as a theoretical counterpart of this practical design principle.

\section{Conclusions}

For many nonlinear channels, common performance measures such as bit error rate, mutual information, and constrained capacity begin to degrade when the transmitted power is increased high enough. The contribution of this paper is to prove that the channel capacity, as defined by Shannon, behaves entirely differently for all channels: Increasing the average power can never degrade the channel capacity, since the adverse effects of high power can always be compensated for by suitably adjusting the source distribution. Until now, the prevalent paradigm in optical information theory has suggested the opposite.

This general result holds regardless of whether the mutual information is maximized over all source distributions with a given average power or all distributions with \emph{at most} the given average power. Indeed, we have shown that the two optimization rules always yield the same channel capacity.

\section*{Acknowledgments}
The author is indebted to 
A.~Alvarado,
L.~Beygi,
G.~Durisi,
T.~Eriksson,
M.~Karlsson,
J.~Karout,
G.~Kramer, and
E.~Telatar
for inspiring discussions and helpful comments on early versions of this manuscript.

\balance %

\end{document}